# Reversible engineering the spin-orbit coupling of monolayer $MoS_2$ via laser irradiation under controlled gas atmospheres

Xilong Liang,[1,2] Chengbing Qin,[1,2*] Yan Gao,[1,2,3] Shuangping Han,[1,2] Guofeng Zhang,[1,2] Ruiyun Chen,[1,2] Jianyong Hu,[1,2] Liantuan Xiao,[1,2*] and Suotang Jia[1,2]

[1] State Key Laboratory of Quantum Optics and Quantum Optics Devices, Institute of Laser Spectroscopy, Shanxi University, Taiyuan, Shanxi 030006, China.

[2] Collaborative Innovation Center of Extreme Optics, Shanxi University, Taiyuan, Shanxi 030006, China.

[3] Department of Physics, Shanxi Datong University, Datong, 037009, China.

[*]Author to whom correspondence should be addressed.

E-mail:

Chengbing Qin, chbqin@sxu.edu.cn

Liantuan Xiao, xlt@sxu.edu.cn

**Abstract**

Monolayer transition metal dichalcogenides (TMDs) with strong spin-orbit coupling (SOC) combined with broken inversion symmetry, leading to a coupling of spin and valley degrees of freedom. These unique features make TMDs highly interesting for potential spintronics and valleytronic applications. However, engineering SOC at room temperature as demand after device fabrication is still a great challenge for their practical applications. Here we reversibly engineer the spin-orbit coupling of monolayer $MoS_2$ by laser irradiation under controlled gas environments, where the spin-orbit coupling has been effectively regulated within 120 meV to 200 meV. Furthermore, the photoluminescence (PL) intensity of B exciton can be invertible manipulation over 2 orders of magnitude. We attribute the engineering of spin-orbit coupling to the reduction of binding energy combined with band renormalization, originating from the enhanced absorption coefficient of monolayer $MoS_2$ under inert gases and subsequent the significantly boosted carrier concentrations. Reflectance contrast spectra during the engineering stage provide unambiguous proof to support our interpretation. Our approach offers a new avenue to actively control the SOC strength in TMDs materials at room temperature and paves the way for designing innovative spintronics devices.

## 1. Introduction

Spin-orbit coupling (SOC), the relativistic interaction between the spin and momentum degrees of freedom of electrons[1], has attracted tremendous attention both for their interesting fundamental physics and potential applications in next-generation devices[2-4]. Essentially, SOC plays a crucial role in condensed-matter physics and accounts for a broad range of fascinating phenomena, such as spin Hall effects[5,6], topological insulation[7,8], Majorana and Weyl fermion[9,10]. On the other side, this interaction also provides a route towards manipulation of electron spins and thus lies at the core of spintronics[11,12]. According to the symmetry dependence, two alternative mechanisms are responsible for the origin of spin-orbit coupling. For the symmetry-independent case, SOC originates from the spin-orbit interaction in atomic orbitals and thus exists in almost all the physics systems[13]. Nonetheless, the symmetry-dependent SOC is only present in crystal structures with broken inversion symmetry, which can be further subdivided into two different types, namely Dresselhaus and Rashba SOC, related to different sources of the electric field asymmetry[14]. Dresselhaus effect is caused by the bulk-induced asymmetry (surviving in the materials without any inversion centers), while the Rashba term is originating from surface- or interface-induced asymmetry (occurring in the systems with broken out-of-plane mirror symmetry)[15]. Generally, these two types of SOC naturally coexist and act on the electron spin as an effective magnetic field[16]. Despite the encouraging progress on the generation and manipulation of SOC in many physical systems (including ultra-cold atoms[17], quantum wells[18], nanowires[19], and organic semiconductors[6]), however, exploring the practical spintronic devices at room temperatures still has proven difficult due to their modest spin-orbit interaction. To this end, physical systems with enhanced SOC are highly desirable[20].

Recent development in atomic-layered transition metal dichalcogenides (TMDs) with chemical formula $MX_2$ (such as $MoS_2$, $MoSe_2$, $WS_2$, and $WSe_2$) demonstrate that[21,22], TMDs are expected to be good candidates for spintronic technologies at room temperatures, due to their giant intrinsic SOC[20,23]. The spin-orbit splittings (charactering the strength of SOC) in TMDs monolayer induced by spin-orbit interaction are several hundred millielectron volts (meV) at the top of the valence band and a few to tens of meV for the bottom conduction band[22,24]. The origin of this dramatic interaction lies in the relative heavy elements in the TMDs materials and the involvement of the transition metal $d$ orbitals[25]. Specifically, the valence band at the $K$ point of the Brillouin zone is primarily composed of the $d_{x2-y2}$ and $d_{xy}$ orbitals of the metal atom[20,26]. This giant splitting defines two different excitons attached to the two spin-split valence subbands, namely A and B excitons, respectively. On the other hand, the conduction band predominantly arises from the $d_{z2}$ orbital of the metal atom, and thus no significant splitting is expected. Nonetheless, a small spin splitting is still survived after considering the role of less strongly coupled chalcogenide orbitals and the higher-order effect of the metal orbital[27-29]. Despite the giant splitting in the valence bands, engineering spin-orbit splitting is still strongly desired, which is an indispensable element of spintronics. Theoretical calculations have predicted that the spin-orbit splitting in TMDs monolayer can be effectively controlled via charge doping[30], biaxial strain[31,32], as well as co-doping of fluorine and group VA elements[33]. Very recently, spin-orbit engineering in $Mo_{(1-x)}W_xSe_2$ alloy monolayers have been experimentally demonstrated, in the view of the critical role of the transition metal $d$ orbitals[34].

Here, we experimentally demonstrated the engineering of the spin-orbit splitting of monolayer $MoS_2$ through laser irradiation under gas-controlled environments. We show that the spin-orbit splitting, *i.e.*, the splitting between A and B excitons, can be enlarged from 152 meV of the as-prepared sample to 200 meV after irradiation with 532 nm continuum-wave (CW) laser under inert gas atmospheres (including $N_2$ and He). The photoluminescence (PL) intensity of B exciton has also been enhanced by more than two orders of magnitude in this stage. We further proved that the enhanced splitting could be reversibly reduced to 120 meV through laser irradiation under air or $O_2$ atmosphere, associated with the recovery of PL intensity of B exciton. We attribute the reversible engineering of spin-orbit splitting and PL modulation to the enhanced absorption coefficient of monolayer $MoS_2$ and subsequent the increase of carriers concentrations, resulting in the significant Auger recombination and pronounced band renormalization. This hypothesis has been confirmed by the reflectance contrast spectra during the engineering of spin-orbit interaction. Our approach provides new opportunities for reversible manipulating SOC in monolayer TMDs after device fabrications at the desired locations, which has important implications in designing future spintronics devices.

## 2. Results and Discussion

For this study, monolayer MoS$_2$ was prepared by traditional chemical vapor deposition (CVD) on the Si/SiO$_2$ substrate (Six Carbon Technology Shenzhen)[35]. Fig. 1a presents the optical image of a typical sample. Isolated single-crystal MoS$_2$ flakes with a triangular shape and edge lengths ranging from 10 to 30 μm can be visualized. We verify the thickness of these flakes to be monolayer by atomic force microscopy (AFM), as presented in Fig. S1. Almost uniform PL intensity (Fig. 1b) further confirms the high quality of the prepared sample. The corresponding PL spectra have been conducted with the excitation of 532 nm continuum-wave (CW) laser (as the arrow shown in Fig. 1c). The spectral profile can be deconvoluted into two components, with the energy of 1.824 eV and 1.976 eV, respectively. Straightforward, the two peaks can be attributed to the A and B excitons of monolayer MoS$_2$, with the spin-orbit splitting of 152 meV. The energy of excitons, the spin-orbit splitting, and the PL intensity ratio between A and B excitons are well consistent with the relevant reports[36-38].

The optical experiments, including laser irradiation, PL spectra measurements, as well as micropatterning, were performed by using a home-built scanning confocal microscope. The schematic diagram can be found in our previous works and electronic supplementary materials (Fig. S2)[39,40]. Specifically, the prepared sample was placed on a motorized three-dimensional piezoelectric ceramics nano stage, which was covered by a polymethyl methacrylate (PMMA) chamber with a control vacuum valve on top of the plane. MoS$_2$ was engineered by a 532 nm CW laser via an objective (Nikon, 100×, NA 0.9), where PL was also collected by the same objective. All the experiments can be mainly divided into two cases. One is engineering monolayer $MoS_2$ with blowing $N_2$ gases (labeled as $N_2$), the other is irradiating the sample blowing the dry air gases (labeled as air).

Figure 1d presents a typical PL trajectory of monolayer $MoS_2$ under laser irradiation against gas cycling. The irradiation and excitation of the $MoS_2$ sample are both performed by the 532 nm CW laser with a power density of 21 MW/cm$^2$. For the pristine sample in air condition, PL is slightly enhanced (roughly twice) under laser irradiation (Fig. S5), which has been reported by our previous work and other groups[39,41,42]. This phenomenon can be understood as the formation of defects and adsorption of $O_2$/$H_2O$ gas molecules, which will result in the depletion of free electrons and the conversion from $A^T$ to $A^0$. Further irradiation in air condition will lead to the damage of crystal and the quenching of PL (Fig. S5). However, further irradiation in $N_2$ atmosphere emerges an entirely new phenomenon, where PL can be significantly enhanced by orders of magnitude (from 16 kcps (thousand counts/s) to 457 kcps) and can maintain the maximum PL without any quenching, as presented in Fig. 1d. Intriguingly, this enhanced PL can be inverted to the initial value in the air atmosphere by switching $N_2$ to air gases. In this case, PL will reduce rapidly and tend to a stable value almost in one minute. Even more intriguingly, this enhancement and reversal can be manipulated circularly merely by switching the $N_2$ and air atmosphere, as presented in Fig. 1d.

To reveal the underlying mechanism, PL trajectory of the first circle is divided into four stages, highlighted by different colors, as shown in Fig. 2a. To explore the reversible proceeding (PL quenching in air atmosphere) more clearly, here we used a relatively low power density (3.5 MW/cm$^2$) to slow down the quenching process. PL spectra of the four stages have been conducted, as presented in Fig. 3b to 3e, respectively. All the spectral profiles can be well deconvoluted into A and B excitons by Voigt functions (typical fitting curves can be found in Fig. S4). The determined photon energies, full width at half maximum (FWHM), integrated intensities, and the corresponding spectral weights for A and B excitons have been presented in Fig. 3f to 3i, respectively (see Fig. S5 for detailed comparison). Now we provide a comprehensive description of the four stages.

(I) In stage I, PL of A exciton is undoubtedly enhanced, which can be attributed to the conversion from $A^T$ to $A^0$, as we discussed above and reported in the previous works[39,41,42]. We also find that the photon energy of B exciton emerges slightly redshift, which possibly originates from the formation of new defects and the reconstruction of the band structure.

(II) In stage II, the FWHM of B exciton dramatically increases from 90 meV to 336 meV. That's to say, the same amount of B exciton has been distributed into a broad energy region when blowing $N_2$. This unusual broadening of FWHM will not further increase with prolonged irradiation times, as shown in stage III. On the other hand, the integrated PL abruptly quenched. From Fig. 2h, we can find that PL intensity of A exciton almost quenches to the background, while that of B exciton presents slight enhancement. The conversion between A and B excitons can also be observed from the variation of their spectral weight (the ratio of the integrated PL intensity of each component to the total intensity), as shown in Fig. 2i. One should emphasize that both stage I and II cannot be restored by switching $N_2$ to air again. We have confirmed that these two stages can be entirely removed by blowing $N_2$ to the pristine sample with several minutes, rather than by experimenting within the air atmosphere as a starting point (Fig. 5).

(III) In stage III, we can find PL intensity of B exciton incredibly enhanced from 242 kcps for the pristine sample to the maximum 25191 kcps, corresponding to more than two orders of magnitude enhancement. Another distinguishing characteristic is that FWHM of B exciton has remained around 320 meV (Fig. 2g), much broader than the initial value. Thus, PL spectra feature an extremely broadband. During this stage, PL of A exciton is also enlarged by 3 times, with a remarkable redshift in its photon energy and broadening in its FWHM.

(IV) After removing $N_2$, photon energies, FWHM, and PL intensities of both A and B excitons will be restored to the beginning of stage III. The abruption of photon energy and FWHM between stages III and IV are probably due to the excitation power switching from 21 $MW/cm^2$ to 3.5 $MW/cm^2$. Further switching the air to $N_2$, PL intensity, as well as the corresponding parameters, can be reversibly engineered, as presented in Fig. 1d.

Beyond the variation of PL intensity and FWHM, photon energy of the two excitons also show dramatic changes. During stage III and IV, A exciton presents successive redshift from 1.826 eV to 1.785 eV (stage III), and then reverts back to 1.825 eV (stage IV). On the other aspect, B exciton shows more complex behavior. Its photon energy shifts from 1.963 eV to 1.996 eV at the initial of stage III, and then constantly softens to 1.979 eV. In stage IV, photon energy of B exciton restores to 1.944 eV, as presented in Fig. 2f. The spin-orbit splitting can be determined by subtracting the photon energy of B from that of A, as shown in Fig. 3. We can find that the splitting increases from 140 meV (at the initial of stage III) to 200 meV (at the end of stage IV), and then recovers to close 140 meV at the end of stage IV. As we expected the spin-orbit splitting during stages III and IV can be invertible engineered when we recurrently switch $N_2$ and air atmosphere. However, the splitting during stages I and II cannot be restored, as we discussed above. This phenomenon indicates that we can reversibly manipulate the spin-orbit coupling of monolayer $MoS_2$ as demand by laser irradiation, with the switching of $N_2$ and air gases.

To optimize engineering and demonstrate the robustness of this approach, we further perform a systematic study. Firstly, we prove that oxygen ($O_2$) plays a vital role in the PL quenching process by changing the air to the pure $O_2$, as shown in Fig. 4a. Note that PL quenches to the minimum value in hundreds of seconds (under 17.5 MW/cm$^2$) when blowing pure air gases. Yet, this process is almost one order of magnitude faster when blowing pure $O_2$ gas. This can be understood as the density of $O_2$ in the pure oxygen atmosphere is five times of that in the air atmosphere (the two most dominant components in air are 21% $O_2$ and 78% $N_2$.). We farther verify that the reversible engineering of the spin-orbit splitting of monolayer $MoS_2$ and the corresponding PL enhancement can be achieved in other inert gases, such as helium (He) in Fig. 4b. PL evolution in He is similar to that in $N_2$ as well, as presented in Fig. S6. It is worth mentioning that both PL enhancing and quenching in He condition are faster than that in $N_2$ atmosphere. To exclude the thermal effect from the CW laser irradiation with high power density, we switch off the laser during the enhancing process in $N_2$ atmosphere, as presented in Fig. 4c. Note that when we switch off the laser for several minutes (even several hours) and hold the $N_2$ atmosphere, the enhanced PL intensity still maintains when we switch on the laser again. If we switch off the laser and stop the $N_2$ blowing, PL emerges a slight quenching and then shows the subsequent enhancement when the $N_2$ blowing is restored. This slight quenching originates from the resident air when the laser is switched on. This result hints that the spin-orbit coupling engineering through our approach is robust and reliable. The modified splitting persists even without laser irradiation and blowing $N_2$ atmosphere (Fig. 4c), manifesting great potential applications of monolayer $MoS_2$ in diverse spintronic and Valleytronic devices.

The power-dependent PL evolutions are also performed to optimize the engineering process. As illustrated in Fig. 4d and 4e, PL enhancement in $N_2$ atmosphere and the subsequent PL quenching in air condition at the laser power density ranging from 3.5 MW/cm$^2$ to 17.5 MW/cm$^2$ have been determined in our experiment. For both processes, the higher the power density, the faster the enhancing and quenching rate. We find that PL enhancement can be empirically fitted by the formula, $I(t) = I_S + (I_0 - I_S) \cdot \{1 + \exp[(t - t_0) \cdot \gamma_\uparrow]\}^{-1}$, where $I_S$ and $I_0$ denote the saturation and initial PL intensity, and $\gamma_\uparrow$ is the rate of the enhancement process. The fit results, as the solid lines shown in Fig. 4d, are in good agreement with experimental data. The determined $\gamma_\uparrow$ as a function of power density is presented in Fig. S7, from which the acceleration of the enhancement process with the increase of power density can be determined. On the contrary, PL quenching processes can be well fitted by a bi-exponential function $I(t) = I_0 + A_1 \cdot e^{-t \cdot \gamma_1} + A_2 \cdot e^{-t \cdot \gamma_2}$ (as the solid lines presented in Fig. 4e). Fascinatingly, the averaged rate of the quenching process, $\gamma_\downarrow=(A_1\gamma_1+A_2\gamma_2)/(A_1+A_2)$, falls exponentially with the increase of laser power, as illustrated in Fig. S7. Although the mechanisms behind the two empirical equations are under debate and further works are needed, these parameters still can provide a guideline to precisely control the spin-orbit coupling and PL intensity of monolayer $MoS_2$.

Now we can declare that the observed optical behaviors of monolayer $MoS_2$ are profoundly different from the phenomena reported in the previous works, where PL enhancing and quenching were mostly attributed to the conversion between $A^T$ and $A^0$[41,42], due to the formation of new defects and the depletion of free electrons[43,44]. However, the maintained maximum PL intensity and reversible features indicate that laser irradiation does not cause further damage to the crystal structure in $N_2$ or He atmosphere, rather than that in air condition[41,42]. To explore the reversible engineering of the spin-orbit coupling and the corresponding PL intensity, we propose a plausible scenario, as presented in Fig. 5. As we discussed above, the valence band (VB) will split into two subbands (A and B) due to the spin-orbital coupling. After laser excitation, electrons will populate at the lowest conduction band (CB), while holes occupy the highest valence band, resulting in the strong emission of A exciton via radiative recombination, as shown in Fig. 5a. Generally, the radiative and non-radiative processes of excitons in monolayer MoS$_2$ can be expressed as $G=An+Bn^2+Cn^3$, where G is the carrier generation/recombination rate[45]. $n$ is the photo-excited carrier (electron/hole) concentrations. A, B, and C represent the non-radiative Shockley-Read-Hall, radiative recombination, and Auger recombination coefficients, respectively. Here Auger recombination represents the annihilation of an exciton (an electron in the conduction band and a hole in A subband) and an electron in B subband, as shown in Fig. 5b. This phonon-assisted Auger recombination will result in the formation of holes in B subband (Fig. 5c). The radiative recombination between the electrons in the conduction band and the holes in B subband brings PL emission of B exciton (Fig. 5d). From the equation, we can find that Auger recombination becomes dominant at high carrier concentrations. In other words, the higher the carrier concentrations, the stronger the emission of B exciton. In our experiment, the excitation power density was maintained for each measurement, which is not responsible for the increase of carrier concentrations. Potential contributions from laser-induced thermal effect have also been excluded by the PL evolution between switching on and off the laser irradiation (Fig. 4c).

Ultimately, we tentatively attribute the increase of carrier concentrations to the enhanced absorption coefficient of monolayer MoS$_2$ under the blowing of inert gases (N$_2$ and He). The enhanced coefficient probably results from the replacement of adsorbed O$_2$ by inert gases, where laser irradiation provides the activated energy to overcome the barrier of the replacement reactions. Considering that the quenching processes are much faster than the enhancing processes under laser irradiation with the same power density, we can speculate that the activated energy of the forward reaction (replacing $O_2$ by inert gases) is higher than that of the backward reaction (replacing inert gases by $O_2$). The elevated carrier densities not only enhance the Auger recombination but also result in screened Coulomb interaction between charge carriers and possible dense electron-hole plasma[46,47]. According to the previous works,[25,48-50] this produces a pronounced many-particle renormalization effect, broaden the linewidth of resonant excitons, as well as reduce both the optical gap and exciton binding energies. The linewidth broadening is in agreement with our results, as shown in Fig. 2g. For the A exciton ($X^A$), the reduction of the optical gap is more pronounced. Hence the optical transition of A exciton is redshift, as we determined in Fig. 2f. Nevertheless, for the B exciton ($X^B$), we speculate that the reduction of binding energy is more significant (as shown in Fig. 5c), resulting in the blueshift of its photon energy. Consequently, the spin-orbit splitting in the conduction band ($\Delta_{SO}^{CB}$) has been enlarged from a few meV to tens of meV. And more importantly, it can be reversibly engineering as demand by an all-optical procedure without any external electric field.

To prove our hypothesis, we further perform the reflectance contrast spectra during stage III and IV, respectively, as shown in Fig. 5e and 5f. Here $\Delta R/R=(R_{MoS2}-R_{Si/SiO2})/R_{Si/SiO2}$, where $R_{MoS2}$ and $R_{Si/SiO2}$ are the reflected light intensity (see methods). At the beginning of stage III, two peaks around 1.88 eV and 2.01 eV can be determined in the reflectance contrast spectra, coinciding with the absorption of A and B excitons reported in the relevant works[36]. With the increase of laser irradiation in $N_2$ atmosphere, the enhancement and broadening of reflectance spectra can be demonstrated, indicating the validity of our hypothesis. The full recovery of the reflectance contrast spectra in stage IV (Fig. 5f) further supports our conclusions.

Although a plausible mechanism has been proposed, further experiments are still needed to clarify the crucial roles of $O_2$ and inert gases, as well as Auger recombination. Firstly, comprehensive calculations, taking carriers density and Auger recombination into account, should be performed to reveal the role of inert gases and $O_2$ in the photoexcited carriers as well as the spin-orbit splitting associating with the variation of carriers density. Secondly, crystal inversion symmetry breaking together with strong SOC leads to coupled spin and valley in TMDs monolayer.[51] Reversible engineering of SOC offers opportunities for controls of spin and valley in these materials, providing a new route to design the high-performance monolayer TMDs-based valleytronics and opto-valleytronics. Furthermore, this coupling results in an energy separation between the spin-allowed and optically active transitions, as well as the spin-forbidden and optically inactive transitions, namely bright and dark excitons.[52-54] The engineering of SOC changes the carrier populations between the two transitions, and thus new evidences and novel physics may be determined by further experiments. At last but not end, the kinetics of the replacement reaction typically dependents upon temperature and the concentration of reactants. Thus, the enhancing and quenching processes may be controlled by changing the temperature of monolayer $MoS_2$, as well as the concentrations of $O_2$, $N_2$, or He. Particularly, the temperature-dependent PL spectra manifest abundant information involving the increase of FWHM and the exciton-phonon scattering (Auger recombination).

## 3. Conclusion

In conclusion, we have reversibly engineered spin-orbit splitting and PL emission of B exciton in monolayer $MoS_2$ by laser irradiation under controlled gas environments (switching between inert gases, such as $N_2$ and He, and air atmosphere). PL spectra evolutions during the reversible manipulation and power-dependent PL behaviors have been conducted and analyzed. The spin-orbit splitting is modified between 120 meV and 200 meV. PL intensity of B exciton has been enhanced by two orders of magnitude. We also show that the reversible and controllable PL spectral features enable explorations of versatile optical applications, such as optical recording (Fig. S8). After excluding the thermal effect originating from the CW laser irradiation with high power, we propose a plausible mechanism to explore this unique PL feature, where the emission of B exciton is attributed to the photon-assisted Auger recombination. We suggest that Auger recombination can be modified during the replacement reactions between inert and air gases. However, the underlying physics remains elusive and needs to study further.

**Acknowledgement**

The authors gratefully acknowledge support from the National Key Research and Development Program of China (Grant No. 2017YFA0304203), Natural Science Foundation of China (Nos. 91950109, 61875109, 61527824, and 61675119), Natural Science Foundation of Shanxi Province (No. 201901D111010(ZD)), PCSIRT (No. IRT_17R70), 111 project (Grant No. D18001), 1331KSC, and PTIT.

**Method**

**Optical measurements**. Schematic of the experimental setup is presented in Fig. S2. A home-built scanning confocal system based on an invert microscope (Nikon, TE2000-U) was used to engineer monolayer $MoS_2$ and take PL spectroscopy, by utilizing a 532 nm CW laser (MSL-III-532, Changchun New Industries Optoelectronics Tech. Co., Ltd.). After extended by a beam expander and filtered by an excitation filter (LL01-532-25, Semrock), the laser was directed by a dichroic mirror (Di01-RE532, Semrock) towards a dry objective (100×, NA 0.9, Nikon) to engineer and excite the monolayer $MoS_2$, which was placed on a motorized three-dimensional piezoelectric ceramics nano stage (200/20SG, Tritor). The laser spot diameter was around 1 μm. PL from $MoS_2$ was collected by the same objective and passed through the dichroic mirror and a notch filter (NF03-532E-25, Semrock). Then, PL was split into two beam by a beam splitter with the ratio of 7 to 3. The weaker one was further filtered spatially by a 100 μm spatial filter and an emission filter (LP03-532RE-25, Semrock) to block the back scattered laser light. PL intensity was detected by a single photon detector (SPCM-AQR-15, PerkinElmer). The stronger beam was sent through a monochromator with the focal length of 30 cm, equipped with a charge-coupled device (DR-316B-LDC, Andor). Monolayer $MoS_2$ was covered by a polymethyl methacrylate (PMMA) chamber with a control vacuum valve on top of the plane (Fig. S3). Dry air, $N_2$, He, and $O_2$ with controlled flow rate were used to blow the sample. All the experiments was performed at room temperature. For the reflectance contrast measurements, a broadband light (DH-2000-BAL, Ocean Optics) was focused onto the sample by the same objective.

**Figure captions**

**Figure-1.**

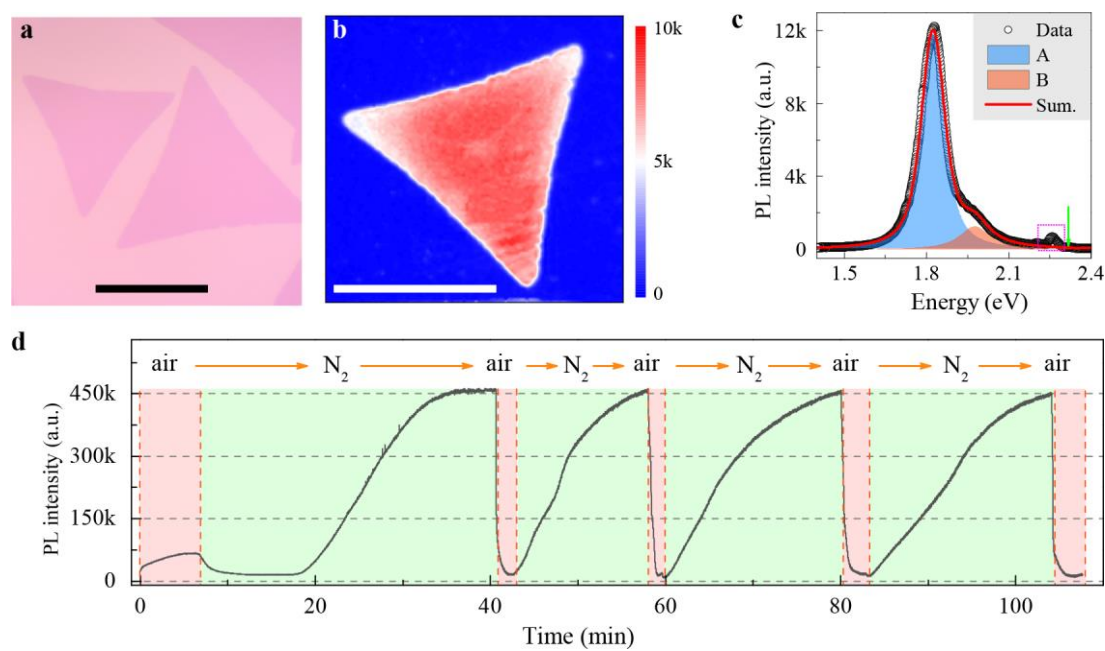

Figure 1. (a) Optical image, (b) photoluminescence (PL) image, and (c) PL spectra of the prepared monolayer $MoS_2$. Scale bar: 20 μm. The spectral profile was deconvoluted by two Voigt functions. The weak peak highlighted by the dashed box is the Raman signal of monolayer $MoS_2$. (d) PL trajectory of monolayer $MoS_2$ under laser irradiation under controlled gas environments. The irradiation and excitation light are both 532 nm CW laser (denoted by the arrow in c). The power density was 21 $MW/cm^2$. Red shades indicate PL behaviors in the air atmosphere, and green shades indicate that in $N_2$ atmosphere. Switching between $N_2$ and Air is achieved by opening or closing the vacuum valve on the top of the chamber. Dashed lines are provided to guide the eyes.

**Figure-2**

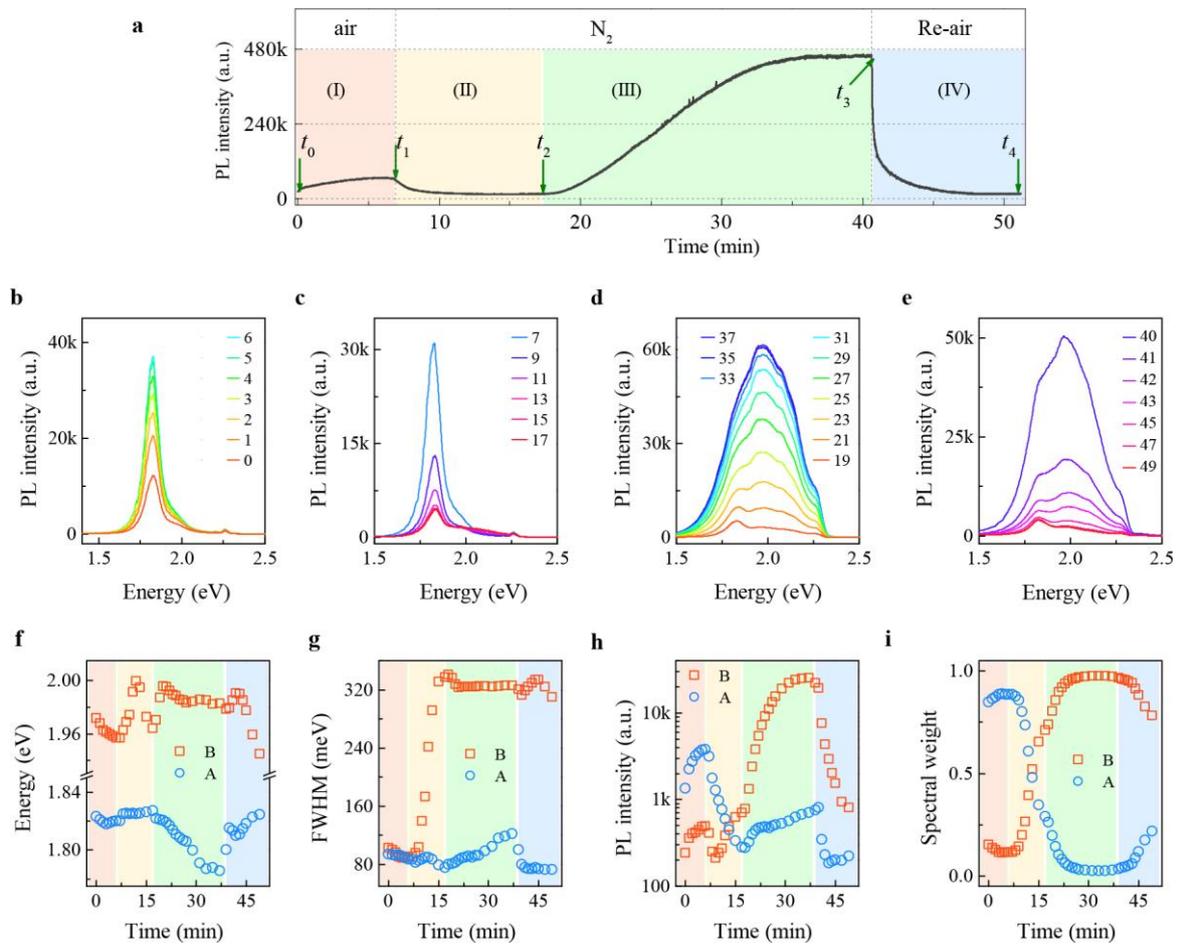

Figure 2. PL evolution of monolayer MoS$_2$ during laser irradiation. (a) The PL trajectory is divided into four stages. The laser power density was 21 MW/cm$^2$ for stage I to III, and 3.5 MW/cm$^2$ for stage IV, respectively. For the convenience of comparison, PL intensity of stage IV has been magnified. For stage I (red) and IV (blue), MoS$_2$ was exposed to the air. For stage II (yellow) and III (green), MoS$_2$ was in N$_2$ atmosphere. $t_0$ to $t_4$ are marked to illustrate the beginning and/or the end of each stage. Dashed lines are provided to guide the eye for comparing PL intensity. (b)-(e) are the spectral profiles of the four stages. The integration time was 3 s; the interval between each spectrum was 1 minute. (f) Photon energies, (g) full width at half maxima (FWHM), (h) the integrated

PL intensity, and (i) spectral weight of A and B excitons as the function of irradiation times. All the parameters are determined by deconvoluting the spectral profiles via Voigt functions.

**Figure-3**

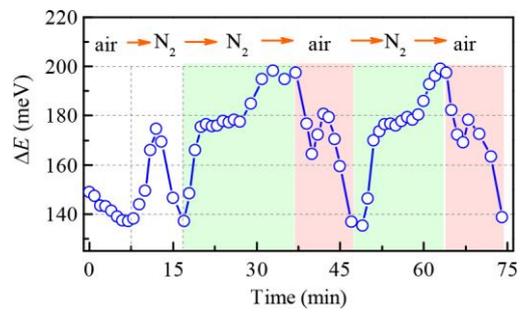

Figure 3. Spin-orbit splitting of monolayer MoS$_2$ as a function of irradiation time during laser irradiation and switching of gases.

**Figure-4**

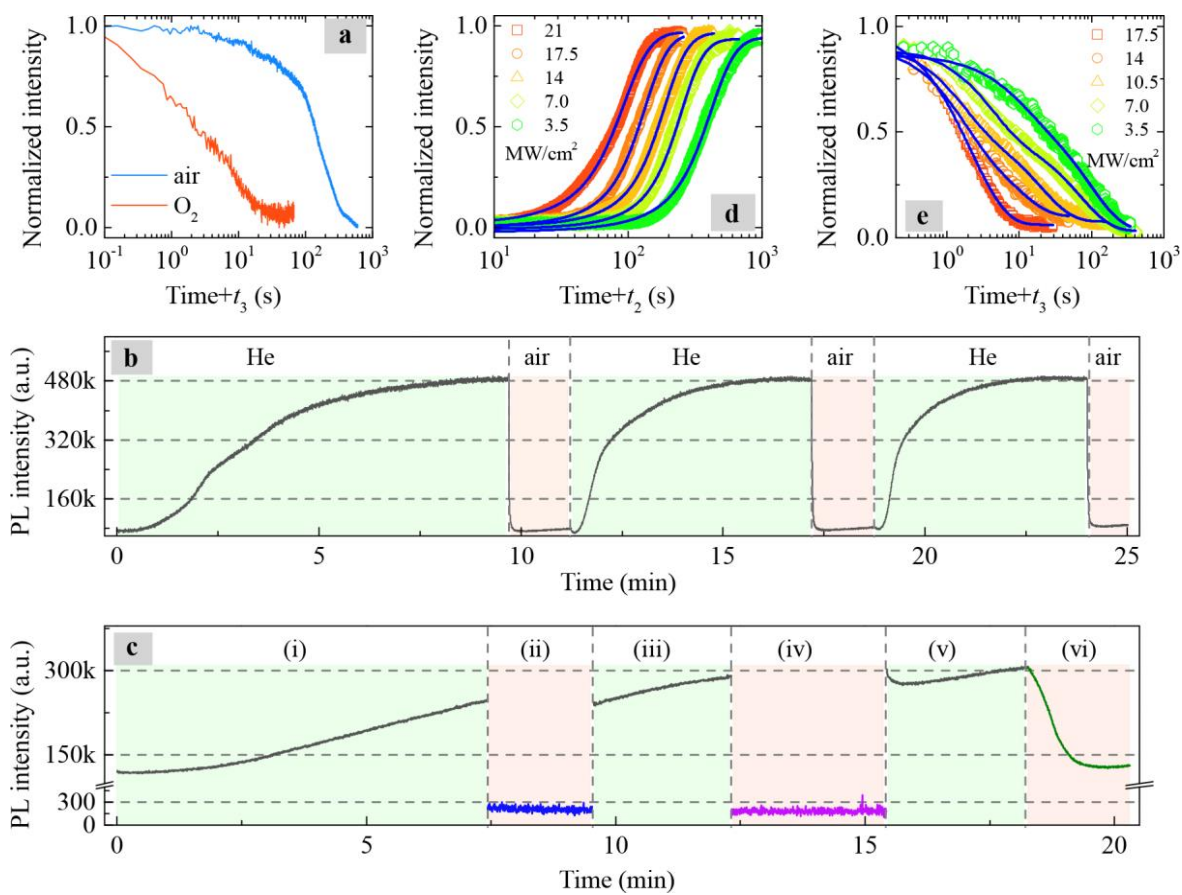

Figure 4. (a) PL evolution under pure oxygen ($O_2$) and air conditions. The irradiation power density was 3 MW/cm$^2$. (b) PL trajectory of monolayer $MoS_2$ against gases cycling between He and air atmosphere. (c) PL trajectory of monolayer $MoS_2$ in different conditions. (i, iii, v) Both laser and $N_2$ are switching on. (ii) Switching off the laser but holding $N_2$ atmosphere. (iv) Both laser and $N_2$ are switching off. (vi) Switching off $N_2$ atmosphere but holding laser excitation. The irradiation power density in b and c was both 21 MW/cm$^2$. (d) PL enhancement in $N_2$ atmosphere, and (e) PL quenching in air atmosphere under laser irradiation with different power densities, respectively. The solid lines in d and e are the fitting results by the corresponding equations discussed in the main context.

**Figure-5**

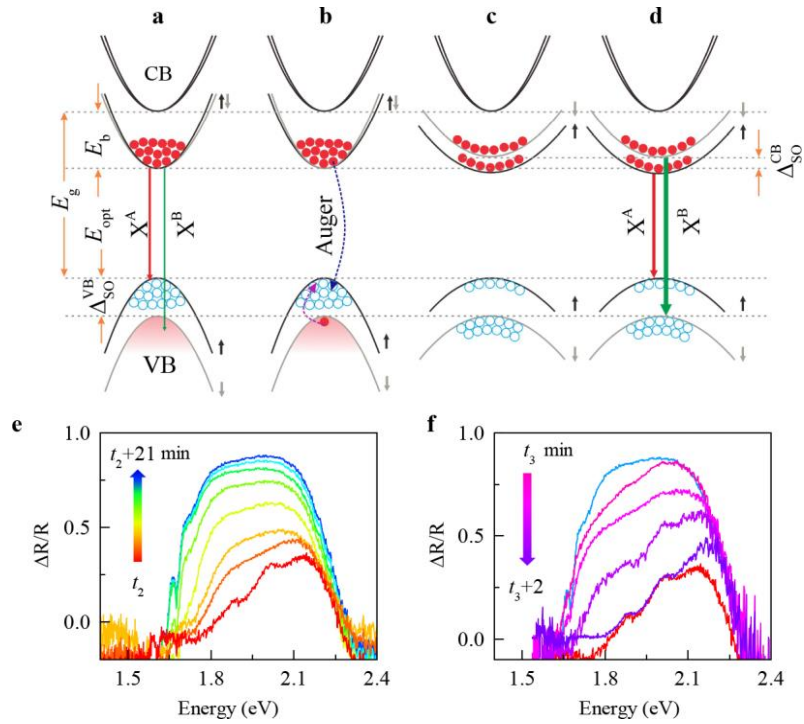

figure 5. Schematic illustration of spin-orbit coupling engineering in monolayer $MoS_2$. Filled (red) and empty (blue) circles indicate the electrons and holes. CB (VB) represents the conduction band (valence band). The spin-up (spin-down) subbands are denoted by the arrows in black (gray) colors. $E_g$, $E_{opt}$, and $E_b$ represent the electronic bandgap, optical bandgap, and the binding energy of excitons, respectively. $\Delta_{SO}^{VB}$ and $\Delta_{SO}^{CB}$ indicate the corresponding spin-orbit splitting of VB and CB. (a) Under low carrier concentrations, the radiative recombination of electrons and holes leads to the PL emission of A exciton ($X^A$). (b) Under high carrier concentrations, Auger recombination becomes dominant, which will annihilate two electrons (one in CB, one in B sub-bands) and a hole (in A sub-band) by the non-radiative process. (c) Auger recombination results in the population of holes in B subband, and subsequent band renormalizations. (d) The radiative recombination of electrons and holes in B sub-band

leads to the PL emission of B exciton. Evolution of reflectance contrast spectra during stage III (e) and stage IV (f) processes shown in Fig. 2. $t_2$ and $t_3$ denote the beginning of the two approaches, as marked in Figure 2a. The time intervals for e and f are 3 min and 20 s, respectively. The highest and lowest curves in F are the end and beginning reflectance spectra in stage III (in e).